\def\~{{$\tilde{\phantom{a}}$}}

\documentclass [12pt] {article}
\usepackage{epsfig}
\usepackage{color}

\def\thebibliography#1{\section{References}\markboth
 {REFERENCES}{REFERENCES}\list
 {[\arabic{enumi}]}{\settowidth\labelwidth{[#1]}\leftmargin\labelwidth
 \advance\leftmargin\labelsep
 \usecounter{enumi}}
 \def\newblock{\hskip .11em plus .33em minus -.07em}
 \sloppy
 \sfcode`\.=1000\relax}
\def\upcite#1{\raise6pt\hbox{\scriptsize
\cite{#1}}}
  \def\lsim{\mathrel {\vcenter {\baselineskip 0pt \kern 0pt
    \hbox{$<$} \kern 0pt \hbox{$\sim$} }}}
    \def\gsim{\mathrel {\vcenter {\baselineskip 0pt \kern 0pt
    \hbox{$>$} \kern 0pt \hbox{$\sim$} }}}

\def\hline{\noalign{\hrule \vskip2pt}}

\def\|{\ifmmode\Vert\else \char`\|\fi}
\ifx\oldzeta\undefined                          
  \let\oldzeta=\zeta                            
  \def\zzeta{{\raise 2pt\hbox{$\oldzeta$}}}     
  \let\zeta=\zzeta                              
\fi

\ifx\oldchi\undefined                           
  \let\oldchi=\chi                              
  \def\cchi{{\raise 2pt\hbox{$\oldchi$}}}       
  \let\chi=\cchi                                
\fi

\def\frac#1#2{{#1 \over #2}}

\def\half{\ifinner {\scriptstyle {1 \over 2}}
   \else {1 \over 2} \fi}

\def\ave#1{\left\langle#1\right\rangle} 

\def\simge{\mathrel{%
   \rlap{\raise 0.511ex \hbox{$>$}}{\lower 0.511ex \hbox{$\sim$}}}}
\def\simle{\mathrel{
   \rlap{\raise 0.511ex \hbox{$<$}}{\lower 0.511ex \hbox{$\sim$}}}}

\def\buildchar#1#2#3{{\null\!                   
   \mathop#1\limits^{#2}_{#3}                   
   \!\null}}                                    
\def\overcirc#1{\buildchar{#1}{\circ}{}}

\def\slashchar#1{\setbox0=\hbox{$#1$}           
   \dimen0=\wd0                                 
   \setbox1=\hbox{/} \dimen1=\wd1               
   \ifdim\dimen0>\dimen1                        
      \rlap{\hbox to \dimen0{\hfil/\hfil}}      
      #1                                        
   \else                                        
      \rlap{\hbox to \dimen1{\hfil$#1$\hfil}}   
      /                                         
   \fi}                                         %

\def\subrightarrow#1{
  \setbox0=\hbox{
    $\displaystyle\mathop{}
    \limits_{#1}$}
  \dimen0=\wd0
  \advance \dimen0 by .5em
  \mathrel{
    \mathop{\hbox to \dimen0{\rightarrowfill}}
       \limits_{#1}}}                           

\def\overlay#1#2{\ifmmode%
\setbox0=\hbox{$#1$}%
\setbox1=\hbox to\wd0{\hss$#2$\hss}\else%
\setbox0=\hbox{#1}%
\setbox1=\hbox to\wd0{\hss#2\hss}\fi%
#1\hskip-\wd0\box1 }

\def\pmb#1{\leavevmode\setbox0=\hbox{#1}%
\kern-.02em\copy0\kern-\wd0
\kern.04em\copy0\kern-\wd0
\kern-.02em\raise.04em\box0 }

\def\vereq#1#2{\lower3pt\vbox{\baselineskip1.5pt \lineskip1.5pt
\ialign{$\m@th#1\hfill##\hfil$\crcr#2\crcr\sim\crcr}}}

\def\tensor#1{\protect\@ontopof{#1}{\leftrightarrow}{1.15}\mathord{\box2}}
\def\overstar#1{\protect\@ontopof{#1}{\ast}{1.15}\mathord{\box2}}
\def\overdots#1{\protect\@ontopof{#1}{\cdots}{1.0}\mathord{\box2}}
\def\overcirc#1{\protect\@ontopof{#1}{\circ}{1.2}\mathord{\box2}}
\def\loarrow#1{\protect\@ontopof{#1}{\leftarrow}{1.15}\mathord{\box2}}
\def\roarrow#1{\protect\@ontopof{#1}{\rightarrow}{1.15}\mathord{\box2}}

\def\@ontopof#1#2#3{%
{\mathchoice
{\@@ontopof{#1}{#2}{#3}\displaystyle\scriptstyle}%
{\@@ontopof{#1}{#2}{#3}\textstyle\scriptstyle}%
{\@@ontopof{#1}{#2}{#3}\scriptstyle\scriptscriptstyle}%
{\@@ontopof{#1}{#2}{#3}\scriptscriptstyle\scriptscriptstyle}%
}%
}

\def\@@ontopof#1#2#3#4#5{%
\setbox0=\hbox{$#4#1$}%
\setbox1=\hbox{$#5#2$}%
\setbox2=\hbox{}\ht2=\ht0 \dp2=\dp0 %
\ifdim\wd0>\wd1 %
\setbox1=\hbox to\wd0{\hss\box1\hss}%
\mathord{\rlap{\raise#3\ht0\box1}\box0}%
\else   %
\setbox1=\hbox to.9\wd1{\hss\box1\hss}%
\setbox0=\hbox to\wd1{\hss$#4\relax#1$\hss}%
\mathord{\rlap{\copy0}\raise#3\ht0\box1}%
\fi
}%

\def\lambdabar{\protect\@lambdabar}
\def\@lambdabar{%
\relax
\bgroup
\def\@tempa{\hbox{\raise.73\ht0
\hbox to0pt{\kern.25\wd0\vrule width.5\wd0
height.1pt depth.1pt\hss}\box0}}%
\mathchoice{\setbox0\hbox{$\displaystyle\lambda$}\@tempa}%
{\setbox0\hbox{$\textstyle\lambda$}\@tempa}%
{\setbox0\hbox{$\scriptstyle\lambda$}\@tempa}%
{\setbox0\hbox{$\scriptscriptstyle\lambda$}\@tempa}%
\egroup
}

\def\corresponds{{\lower.2ex\hbox{=}}{\rm\kern-.75em^\triangle}}
\def\succsim{\succ\kern-.9em_\sim\kern.3em}
\def\precsim{\prec\kern-1em_\sim\kern.3em}
\def\slantfrac#1#2{\kern1em^{#1}\kern-.3em/\kern-.1em_{#2}}

\begin{document}

\begin{center}
{\Large\bf The Laser Driven Vacuum Photodiode}
\\

\medskip

Kirk T.~McDonald
\\
{\sl Joseph Henry Laboratories, Princeton University, Princeton, NJ 08544}
\\
(Sept.~26, 1986)
\end{center}

\section{Problem}

A vacuum photodiode is constructed in the form of a parallel plate capacitor
with plate separation $d$.  A battery maintains constant
potential $V$ between the plates.
A short laser pulse illuminates that cathode at time $t = 0$ with energy
sufficient to liberate all of the surface charge density.  This charge moves
across the capacitor gap as a sheet until it is collected at the anode at
time $T$.  Then another laser pulse strikes the cathode, and the cycle
repeats.

Estimate the average current density $\ave{j}$ that flows onto the anode from
the battery,
ignoring the recharing of the cathode as the charge sheet moves away.
Then calculate the current density and its time average when this effect is
included. 

Compare with Child's Law for steady current flow.

You may suppose that the laser photon energy is equal to the work function
of the cathode, so the electrons leave the cathode with zero velocity.

\section{Solution}

The initial electric field in the capacitor is ${\bf E} = -V/d\hat{\bf x}$,
where the $x$ axis points from the cathode at $x = 0$ to the anode.  The
initial surface charge density on the cathode is (in Gaussian units)
\begin{equation}
\sigma = E/4 \pi = - V/4 \pi d.
\label{s1.6}
\end{equation}
The laser liberates this charge density at $t = 0$.

The average current density that flows onto the anode from the battery is
\begin{equation}
\ave{j} = - {\sigma \over T} = {V \over 4 \pi d T}\, ,
\label{s1.6a}
\end{equation}
where $T$ is the transit time of the charge across the gap $d$.
We first estimate $T$ by ignoring the effect of the recharging of the
cathode as the charge sheet moves away from it.  In this approximation,
the field on the charge sheet is always $E = -V/d$, so the acceleration of
an electron is $a = -eD/m =eV/dm$, where $e$ and $m$ are the magnitudes of
the charge and mass of the electron, respectively.
The time to travel distance $d$ is
$T = \sqrt{2 d/a} = \sqrt{2 d^2 m /eV}$. Hence,
\begin{equation}
\ave{j} =  {V^{3/2} \over 8 \pi d^2} \sqrt{2 e \over m}.
\label{s1.6b}
\end{equation}
This is quite close to Child's Law for a thermionic diode,
\begin{equation}
j_{\rm steady} = {V^{3/2} \over 9 \pi d^2} \sqrt{2 e \over m}.
\label{s1.6c}
\end{equation}

We now make a detailed calculation, including the effect of the recharging
of the cathode, which will reduce the average current density somewhat.

At some time $t$, the charge sheet
 is at distance $x(t)$ from the cathode, and
the anode and cathode have charge densities $\sigma_A$ and $\sigma_C$,
respectively.  All the field lines that leave the anode terminate on
either the charge sheet or on the cathode, so
\begin{equation}
\sigma + \sigma_C = - \sigma_A,
\label{s1.7}
\end{equation} 
where $\sigma_A$ and $\sigma_C$ are the charge densities on the anode and
cathode, respectively.
The the electric field strength in the region I between the 
anode and the charge sheet is
\begin{equation}
E_I = - 4 \pi \sigma_A,
\label{s1.8}
\end{equation}
and that in region II between the charge sheet and the cathode is
\begin{equation}
E_{II} = 4 \pi \sigma_C.
\label{s1.9}
\end{equation}
The voltage between the capacitor plates is therefore,
\begin{equation}
V = - E_I (d - x) - E_{II} x = 4 \pi \sigma_A d  - V {x \over d}\, ,
\label{s1.10}
\end{equation}
using (\ref{s1.6}) and (\ref{s1.7}-\ref{s1.9}), and taking the cathode to
be at ground potential.  Thus,
\begin{equation}
\sigma_A = {V \over 4 \pi d} \left( 1 + {x \over d} \right), \qquad
\sigma_C = - {V x \over 4 \pi d^2}\, ,
\label{s1.11}
\end{equation}
and the current density flowing onto the anode is 
\begin{equation}
j(t) = \dot \sigma_A = {V \dot x \over 4 \pi d^2}.
\label{s1.12}
\end{equation}
This differs from the average current density (\ref{s1.6a}) in that
$\dot x/d \neq T$, since $\dot x$ varies with time.

To find the velocity $\dot x$ of the charge sheet, we consider the force
on it, which is due to the field set up by charge densities on the anode
and cathode,
\begin{equation}
E_{{\rm on}\ \sigma} = 2 \pi(-\sigma_A + \sigma_C) 
= -{V \over 2d} \left( 1 + {2x \over d} \right).
\label{s1.13}
\end{equation}
The equation of motion of an electron in the charge sheet is
\begin{equation}
m \ddot x = -e E_{{\rm on}\ \sigma}  
= {eV \over 2 d} \left( 1 + {2x \over d} \right),
\label{s1.14}
\end{equation}
or
\begin{equation}
\ddot x - {e V \over m d^2} x = {e V \over 2 m d}.
\label{s1.15}
\end{equation}
With the initial conditions that the electrons start from rest,
$x(0) = 0 = \dot x(0)$, we readily find that
\begin{equation}
x(t) = {d \over 2} (\cosh kt - 1),
\label{s1.16}
\end{equation}
where
\begin{equation}
k = \sqrt{e V \over m d^2}.
\label{s1.17}
\end{equation}
The charge sheet reaches the anode at time 
\begin{equation}
T = {1 \over k} \cosh^{-1} 3.
\label{s1.18}
\end{equation} 

The average current density is, using (\ref{s1.6a}) and (\ref{s1.18}),
\begin{equation}
\ave{j}= {V \over 4 \pi d T}
={V^{3/2} \over 4 \pi \cosh^{-1}(3)\ d^2} 
\sqrt{e \over m}
= {V^{3/2} \over 9.97\ \pi d^2} \sqrt{2 e \over m}.
\label{s1.21}
\end{equation}

The electron velocity is
\begin{equation}
\dot x = {d k \over d} \sinh kt,
\label{s1.19} \end{equation}
so the time dependence of the current density (\ref{s1.12}) is
\begin{equation}
j(t) = { 1 \over 8 \pi} {V^{3/2} \over d^2} \sqrt{e \over m} \sinh kt
\qquad (0 < t < T).
\label{s1.20}
\end{equation}

A device that incorporates a laser driven photocathode is the laser triggered
rf gun \cite{bnlgun}.


\begin{thebibliography}{99}
\bibitem{bnlgun}
K.T.~McDonald,
{\sl Design of the Laser-Driven RF Electron Gun for the BNL Accelerator 
Test Facility}, 
IEEE Trans.\ Electron Devices, {\bf 35}, 2052-2059 (1988).

\end{thebibliography}
\end{document}